\documentclass[12pt]{article}
\usepackage[latin1]{inputenc}
\usepackage[english]{babel}
\usepackage{textcomp}
\usepackage{mathptmx}
\usepackage{helvet}
\usepackage{courier}
\usepackage{pstricks}
\usepackage{makeidx}

\makeindex

\def\us{\char`\_}

\begin{document}

\author{José del Carmen Rodríguez Santamaría }
\title{ The genome is software and evolution is a software developer}

\date{}

\maketitle

The genome is software because it a set of verbal instructions for a programmable computer, the ribosome. The theory of evolution now reads:   evolution is the software developer responsible for the existence of the genome. We claim that this  setting, whose official name is genetic programming, is necessary and sufficient to discuss all important questions about evolution.    A great effort has been made to pass from wording to science, i.e., from naive theories to robust models to predictions to testing for falsification.

\section{Basic concepts}

 We, human beings, use languages to communicate one with another. A  \textbf{language} is composed of verbal instructions, a code of interpretation and a semantic content. A   \textbf{verbal instruction} is a string of sounds or letters that belong in an   \textbf{alphabet}.  Verbal instructions must be deciphered according to a   \textbf{code of interpretation}, which appears, say, in a dictionary. The deciphered   verbal instructions convey information about the world or about actions that must be undertaken and it is in this way that they acquire  \textbf{semantic content}.

A \textbf{computer }  is a devise that has the potential ability of executing verbal instructions. A \textbf{program}   is a set of verbal instruction, written in a specific   \textbf{programming language} that can be interpreted and executed by a computer. \textbf{Code} is a set of instructions that eventually contain correctly written programs.    \textbf{Software} is a set of programs to execute a task. \textbf{ Hardware} is the wiring that converts programs into specific actions. The core of the hardware is the microprocessor that contains the central processing unit, which can understand verbal instructions that are written in machine language, which is composed of words whose alphabet is just $\{0, 1\}$. An  \textbf{operating system} is a set of programs specifically targeted to serve the needs of communication between  human users and the hardware: it contains the know-how of operating the computer to execute a verbal instruction that commands to, say, print a letter on the console; it also produces feedback from the hardware, say, it could communicate that there is no network connected to the computer. Simple microprocessors are not enabled to admit operating systems apart than the one contained in its machine language. A \textbf{robot}  is a machine that is operated by a computer that has been programmed to have certain independence.

\

A  \textbf{designer} is a person that is committed to the design of software with a specific function, which is defined before beginning but that might be modified along the process. 

\

 Let us define \textbf{SOFT} as the problem of designing software. Soft has a prelude: if one works in a company and the program must be done for a client, he or she must say what function  the program must fulfill. If this is officially done, the function is transcribed to a document that is called \textit{spec}.

 More specifically, spec contains the requirements that a system must fulfill: \textit{on this input, and under given parameters, the system must  answer like this but with this other input the system must be answer like that}. For instance, if the user types the name of a client, the system must display desired information such as  photo, telephone number and address but the type of music he or she prefers must be displayed only if that is asked in advance.   All this might be encoded as a huge table that    assigns   binary strings to binary strings. The task for the developer is to find an \textbf{algorithm} or procedure  that  reproduces the correct assignations and to encode it into a convenient programming language. 
 
  A \textbf{bug} is any error against a goal in programming. It is absolutely sure that the first trial of a code is inviable.   But if one manages to produce a viable code, most surely it will not do what one was so eagerly expecting. The process of adjusting the code to the predefined purpose is called  \textbf{debugging}.
 
From the linguistic stand, spec is no more than an extended   proposition, whose overall form is \textit{do this but not that}. The sentence \textit{do this} plays the role of a verb and represents elementary executing actions, while the sentence \textit{but not that}   plays the role of an adverb and represents all the possible regulations that  actions may include.
 
 \
 
In contrast to a software designer, which is a person with a goal, we also have the concept of  \textbf{software developer}, which is any process or entity that produces software.  The    point  to be consider in this article   is that \textit{the genome is software and that evolution is a software developer whose possibilities are worth being studied both to explain the genome and to be used in industrial software design}.

\

 We highlight the fact that the concept of bug has a meaning in software design but in principle  it lacks any meaning in software developing. The reason is that a bug is a fault against a goal, a project, a predefined function, while software developing may run without predefined goals by mere diffusion. Nevertheless, in hindsight, we can speak of bugs or errors\textbf{} of evolution (see below).

\section{The genome is software}

The \textbf{genome}   contains, apart from other things, a set of     instructions to program ribosomes to build specified proteins (Darnell et al, 1986; The National Health Museum, 2007). From the linguistic stand, the genome is indeed software: it is based on an alphabet, it conveys instructions to be executed, it has a code and obeys a given semantics:

\subsection{The alphabet }

For DNA, the alphabet is $\{ A, T, C, G \}$. For RNA, it is $\{ A, U, C, G \}$. Anyway, there are other molecules that also count, say, the TT dinucleotide, which is formed by illuminating with ultraviolet rays the DNA. There are also molecules that can be inserted, maybe by van der Wans interactions, into a DNA string producing lethal consequences.

\subsection{The code and the semantics}

The code that allows the deciphering of the verbal instructions contained in the genome is the  \textbf{genetic code}, a dictionary that contains the correspondence among  \textbf{codons} ( a triplet of nucleotides)  and amino acids (Leroy and Galas, 2003). This dictionary is written in the t-RNAs: these molecules have in one extreme an anti- codon (complementary to a codon) and in the other the corresponding amino acid. Strictly speaking, the code is written not in the t-RNA's but on the enzymes that synthesize them, the aminoacyl-tRNA synthetases (Darnet el al, 1986). Amino acids are assembled together by the ribosomes into proteins, which   serve as structural items or as enzymes to catalyze chemical reactions  that   enable life and  surviving. Thus,   \textbf{surviving} is the feature that allows the genome to acquire a \textbf{semantic content}: a viable string has a meaning but an inviable one does not. A \textbf{cell} is a robotic entity: it is directed by  software and has a lot of independence.

One can distinguish two type of genes: some encode for specific actions while other encode  for regulations of  those actions. A gene of the first type is called \textbf{structural}, of the second is   \textbf{regulative}. Example: rhodopsin is the protein that responds for the perception of light in our eyes. There is a gene that encodes for that protein, it is structural. But when one just passes from light to darkness,   one gets blinded but in some  few minutes one can see again: the change from light to darkness triggered a regulative response in which more rhodopsin was synthesized. Such a response is enabled by  appropriate regulative genes.

\subsection{The ribosome is a programmable computer}

Portions of  DNA are transcribed into RNA, whose codons are interpreted by t-RNA's and the ribosome to build proteins. The whole ensemble functions like a programmable computer: one can order the synthesis of whatever protein one wants. The   ribosomes are the hardware and the operational system is the code properly, in the t-RNA's (Goldenfeld and Woese, 2007). 

Thinking that  the ribosome alone is responsible for the execution of natural computing programs could be misleading: the genome  is as a library with all   administrative duties of a library of books for human use and the material received by the ribosome is a product of a process of elaboration made by a set of enzymes involved in opening frames, DNA transcription to RNA, RNA maturation, and this work continues even after the proteins   exit the ribosome. All these operations could be understood as supportive,  syntactic and grammatical rules.

\subsection{The concept of software  is magnificent}

Let us see why   computers are very different  than the rest of physical systems.

Large physical systems, stones and cars, have a tendency to follow in the same direction and with the same velocity unless external forces, say friction, exert some action over them. In general, the physical word is in first instance characterized for its tendency to behave just in the form it behaved in the past.

By contrast to matter, we are living beings that, although subjected to physics, can manage situations to behave as free entities,  doing what we plan to do. We are men and   women that are so conscious of our freedom that in most cultures we are considered to be responsible for  our words and deeds. What physical properties enable the existence of our freedom? It is  a robotic facility possessed by living beings, which is  a system of computation to calculate decisions plus a system to execute received commands. That facility enables free movement.

Artificial robots are beginning to invade our lives, so we are completely sure that they have nothing else apart from highly organized matter. So, we ask: if a robot is only mater, how does it acquire freedom from the inertial behavior?

The answer is this: a robot never ceases to obey physical laws, it is precisely because of them that robots function. The physical laws enable robotic freedom as follows. In first instance, some physical systems	 may behave erratically, say, the trajectory of a  particle that travels inside the water of a turbulent river cannot be predicted. The behavior of such a particle is so distinct than that of a stone that science created  a whole world for its a study,  \textbf{chaos}, that happens when a minor change in the initial condition may cause tremendous changes in the trajectory, so much that these systems being deterministic are best studied by statistical means, as if they were random systems (Baranger, 2010).  In fact, chaos is used in modern technology to simulate  random sources: chaos is the deterministic image of randomness. 

Chaos is also at the essence of computer units:  the form as a software developer perceives the  chaotic nature of the world of computer programs is through \textbf{bugs}: a minor change in the code  may cause tremendous changes in the form of operation or may  lead to death.  But we know that a chaotic personality is not desirable for a human being. The matter is that humans are free but controllable. So is software.

Thus, most humans and all living creatures and all robots and all computers enjoy two properties: they are almost chaotic and also are controllable.  That is easy to say but it is very difficult to implement and that is why computers were only recently built. At the heart of computers, we find   systems that may answer all or nothing  to a given input and that can be as simple as a diode or as a transistor. Elements that answer all or nothing to an input are example of   \textbf{digital} systems. DNA bases are digital in the sense that they recognize their complementary partners, C with G and A with T, and so they answer yes (you are my partner) or not (you are not my partner) to an incoming molecule (Hood and galas, 2003). Combining simple elements, one can achieve great levels of complexity, such as those  we see in modern computers, in any programming language and in the genome.

So, software is what makes the difference between stones and free will of action. Most probably, there is no higher exaltation of the power of software than that presented by Apostle John. Of course, he does not uses the word software, instead he uses \emph{word}, the mean that human beings use   to think, to make calculations,  to express   decisions and to teach, i.e.  to program other people. He opens his Gospel with the next words:

$E\nu \ \ \alpha\rho\chi\eta \  \ \eta \nu \ \ o \ \ \lambda  o \gamma o \varsigma, \ \ \kappa \alpha \iota \ \ o \ \ \lambda  o \gamma o \varsigma \ \ \eta \nu \ \ \pi \rho o \varsigma \ \ \tau o \nu \ \ \Theta  \epsilon o \nu,$

$  \kappa \alpha \iota \ \ \Theta  \epsilon o \varsigma \ \  \eta \nu \ \ o \ \ \lambda  o \gamma o \varsigma.$

This sentence can be read as follows:

En arkhe en  o logos, kai o logos en  pros ton Theon, kai Theos en o logos.

A literal translation reads:

\emph{At the beginning was the Word, and the Word was with God, and God was  the Word}.

The official translation is:

\emph{In the beginning was the Word, and the Word was with God, and the Word was God} (The King James Version, Cambridge 1769).

These words were written for people living two thousand years ago under the power of the Roman Empire. Can we understand what that message is about? Yes, of course. Let us consider  a stone that is released from a tower, it falls down inexorably and continues movement until it collides with the earth. The stone contrasts with a horse that somehow can be guided using love an authority. By comparison, a small daughter obeys with pleasure precise commands in the measure of its possibilities. Nevertheless, there is nothing like a legion: if the commander says to one man, go, he goes. And if he says to another, come, he comes. Or to a servant, do this, and he does it. And one can add command after command to produce a plan of action  and simple say: \textit{vine, vide, vinci}. 

As we see, the lines of authority function in the very same form as  algorithms do and also include  high levels of \textbf{recursion}, in which one  can package a very long list of instructions  into a directive or  a established protocol. One can convert desires in commands and these in actions if only one is able to build a much surely very complex machinery that can interpret commands and execute them. In this game, interpretation and execution go hand in hand, they have no meaning in separate terms but together convert the ruler into a sort of god. In fact, the Emperor was given the rank of god all throughout the Roma Empire and he was worshiped as Zeus or Diana. 

The notion of algorithm is not restricted to lines of authority and any person  is well acquainted with intuitive notions of algorithms and their power. Examples:

\begin{enumerate}
\item Cooking receipts, that could be very complex.
\item Instruction to build houses and buildings.
\item  Industrial procedures as those to make  tools for chasing and/or war.
\item  Agricultural know-how that includes golden secrets, say, the  seed of wheat is sowed at the end of the autumn and it remains under the earth the whole winter so that it could sprout up as soon as the spring begins.
\end{enumerate}

\subsection{Justice has been made}

 Given that  the genome is   the natural example of software, \textbf{the  evolutionary theory} now reads:    evolution  is the software developer responsible for the existence of the genome. 
 
 We hope that our new insight  at last has made  justice to all biologists that instinctively consider that life is the summit of marvels. The reason is that  software is possibly the most sophisticated and difficult high precision tool to devise because minor variations produce no function else a very different function than the searched one.

The extreme difficulty of software design makes one to look at the evolutionary theory with suspicion:  given that one  suffers so much to compose a functional piece of code, how can one believe that a process that has no intelligence could be able of concocting so great diversity of genomes that contain instructions for the most sophisticated products in the universe? This intuitive appreciation is important in the measure that it could be transcribed into a scientific program of research. In this regard, our approach is as follows: 

If the aforementioned subjective complaint is correct, there must be observable facts with the power to  prompt every scientist to conclude that the evolutionary theory has been falsified, i.e. that very important predictions of the theory are not fulfilled by nature or by experiment. So, what are the predictions and where are their falsification?

Guided by the human experience, as in academy as in industry and corroborated by every practitioner,  let us emphasize that perfection in software design is only statistical and begins to appear after a very tortuous path that includes the correction of many bugs with varying degrees of fatality. This experimental generalization is so infallible that we promote it to a law for software design, the \textbf{bug law}: there is no functional  software without mountains of bugs that are followed by corrections that generate more bugs. This law  is valid by construction for the software that is  designed by human beings.

While the developer views a developing path paved in bugs, the vision of the user is very different:  the user sees an evolution that goes from not a bad product to a good project to a better one to excellent quality. When projects are very complex, with every kind of multifunctionality, the user also shares the feeling of dissatisfaction caused by half cooked products.

\stepcounter{figure}
\bigskip
\begin{center}
\psset{unit=0.05cm}
\begin{pspicture}(0,0)(160,103)
\psline(20,10)(30,10)(35,50)(38,60)(40,60)(40,10)
(60,10)(60,20)(65,20)(65,10)(70,10)(70,30)(75,30)
(75,10)(80,10)(80,40)(85,40)(85,10)(90,10)(90,70)
(95,70)(95,10)(100,10)(100,80)(105,85)(105,80)
(110,80)(110,85)(115,90)(120,90)(120,95)(125,95)
(125,90)(130,90)(130,100)(135,100)(145,80)

\psellipse(35,60)(10,-5)
\psellipse(122.5,95)(5.5,-3)
\psellipse(132.5,100)(5.5,-3)

\psline{->}(20,10)(160,10)

\psline{->}(20,0)(20,102)

\rput(120,0){\emph{Coordinate along a developing path}}
\rput(40,108){\emph{Subjective feeling of perfection}}
\rput*(20,90){\emph{Releases perceived by the user}}
\psline(44,83)(39,67)
\psline(68,89)(113,94)
\psline(68,93)(124,100)
\end{pspicture}

\bigskip

\end{center}

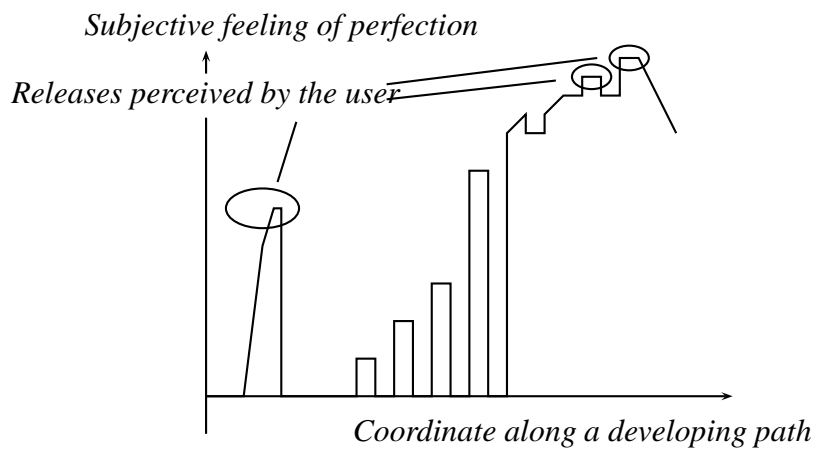
\textit{Figure \thefigure. Evolution of software is mandatory but the perception of this process is different for the developer team (broken line) than for the user (Elliptical enclosures).    }
\stepcounter{figure}

\

As a consequence, commercial software always leaves clear fossil evidence of  products that are improved over generations and that may remain \textbf{half cooked}, unfinished and with low quality. For instance, Microsoft  Windows has had many versions that, in general, have acquired more quality over successive releases but, anyway, Windows Vista might be blamed as a half cooked product because of its deadlocks, crashes and many incompatibility problems.   These type of troubles  are  common place in the deal. 

Our prediction extends the bug law to natural evolution: given that evolution is a developer, the   path followed by natural evolution must be filled in every sort of bugs,  half cooked products as in the fossil record as in modern populations. Where is the falsification of this prediction? It is here:

We see no other thing in living beings apart from extreme perfection. Therefore, the evolution depicted by modern science corresponds to the description of an evolutionary process as reported by the  user. As a consequence,  both, science and nature, lack a  mandatory description of evolution toward perfection that corresponds to the developer and that must be  filled in pitfalls of every sort.  

\

We have managed to pass from a subjective discomfort to operational propositions of scientific value. That was a great lap. Nevertheless, we hurried a lot an so we were unable to perceive that we have applied to a material process results that are valid for human beings. Is that correct? To answer this question, we must make a large detour in which we try to unveil how is the evolution of perfection of complex traits.  The first step is, of course, to understand better what we mean when we say that:
 
\section{Evolution is a software developer}

Since we find mutation at the core of  evolution,  the evolutionary theory implies  that mutation is a software developer or that with mutations we can generate software. Does this make sense? Yes. To see this, we need to notice before everything else that words can convey computing information. To be sure, modern programming languages exist just because words have intrinsic computing power. There are various forms to see this but the most clear is to recall  that a modern computing language is an interface between a human being, who operates with words, and a computer that operates with zeros and ones.  So, if one gives a glance to a program in any (high level)  programming language, one sees words and no more than words that with some training can be clearly understood. Therefore, words can convey information to make the very same calculations that are done by any computer. 

Let us consider now the proposal that one can generate computer programs with mutations, i.e. by modification of extant texts. As dangerous as this might look  this is obvious: one can use mutations to go from a text with no interest, say,   \textit{aaaaaaaaa}, to whatever other text and in particular to the text of anyone computing program. By the way, human writers  use only three types of mutations to go from a blank document to finely refined computer programs: insertion, deletion and translocation (copy from here and paste in there). 

The point is that chance can produce concatenations of mutations that eventually could produce over, say,  a blank document, outputs that serve a purpose under appropriate circumstances.  We consider that this assertion is obvious and deserves no discussion. Now, the point of the evolutionary theory is that chance is maximally  profited in an evolutionary environment with  mutation and selection. It is our duty to show how this  theory can be  calculated with the aim of making serious predictions that next must be contrasted with nature to see whether or not these are fulfilled. 
 
 \
 
At the meantime, let us see a concrete example that shows that  with mutations one can generate instructions to calculate and specific function.  To that aim, let us consider the   next sequence of strings:
 
 $AG$
 
 $AGAG $
 
 $AGAGAG$
 
 $.. ..$
 
 $AGAG....AGAG$

We see here a recursive application of the   instruction:

 \textit{Insert AG to the end of the previous string}.   
 
Now, we can interpret this sequence as follows:
 
$AG \rightarrow 0.10 \rightarrow (\frac{1}{10})^1  $
 
 $ AGAG \rightarrow 0.1010  = 0.10 + 0.0010 = (\frac{1}{10})^1+(\frac{1}{10})^3 )$
 
 $AGAGAG \rightarrow 0.101010 = 0.10 + 0.0010 + 0.000010 = (\frac{1}{10})^1+(\frac{1}{10})^3 + (\frac{1}{10})^5 $
 
 $.. ..$
 
 $AGAG....AGAG \rightarrow 1010..1010 = \sum^n_1 (\frac{1}{10})^{2k-1}$

 Thus, our insertion rule generates a chain of instructions that compute the series $\sum^\infty_1 (\frac{1}{10})^{2k-1}$    more and more approximated after each computation if  A is interpreted as 1 and G as 0. This example can be generalized by saying that with words and with mutations   one can generate programs, routines, methods to calculate functions. Functions that can be calculated by means of procedures of this sort are called \textbf{recursive functions} (Brookshear, 1989; Lewin, 2010). When strings of symbols and mutations are used to create and execute  computations, we speak of \textbf{mutation machines}.
 
 \

To be fair, our mutation machines seem to be more in our mind than in the cell. Strictly speaking, the ribosome with the support of the whole cell runs the programs consigned in the DNA. What are these programs for? They direct the synthesis of proteins and enzymes. But, where are the recursive functions? To understand where are they, let us consider a muscular cell, containing actin and myosin. They are encoded by the genome. These polypeptides form part of an ensemble that  executes in marvelous way the function of converting   free physic-chemical energy into  movement. It is the similar of the motor of a car, only that the muscular unit is molecular and not macroscopic. Now, the idea  of a motor is not a common one: ancient Greeks knew all that one must know to built a vapour motor. Moreover,  it is not easy to make a good motor: we are still battling to make a  motor as effective as the cellular ones (He et al, 2000; Fueleconomy, 2010) and on this may depend the future of the planet because of the global warming effect.

We see that the genome contains the solution to terrible mathematical problems that we try to solve with the use of models, recursive functions and computers. So, the recursive functions are there but in concealed form because the genome do not solve abstract problems but real ones and without modeling. 

\

We tacitly are assuming that the complexity of a problem is intrinsic and that the problems that evolution must solve in vivo are roughly as complex as those that we generate with fair models and that try to solve with recursive functions in silico. This very delicate question
can be saved for a while if we formulate another more tractable one: can we    build or simulate mutation machines that can solve the most terrible mathematical problems?  This question is nevertheless so extreme that the answer is most probably negative, so we prefer to reformulate it in more pragmatic terms:  can we    build or simulate mutation machines that can solve the most terrible mathematical problems that can be solved with other computers? We urgently need an affirmative answer to this question  otherwise it would be inadequate for our honor to propose that evolution could be the developer of the genome and of all the marvels that it encodes for.  The official  answer to the last question is found along the next lines.

\section{All  computing paradigms are equivalent}

We have  two very distinct models or paradigms of computation, numbers and words. It is important to pick up the paradigm that is the best, say, with more expressive power or with more velocity. The answer to this problem is that a computer calculates  recursive functions and all computers   calculate in the same regime the very same class of recursive functions.  A given paradigm can be a bit better than other for a specific problem but no paradigm is the best for everything.

The equivalence of  all known  paradigms is proved constructively and in reference to a Turing machine (Caicedo, 1990;  Cousineau, 2000) that is the official ideal computer. The corresponding theory can be found in various versions. One, appropriate for researchers in molecular biology, can be found in  Rothemund (1996). The aim of research is to design and build artificial  mutation machines, whose first exponent  was  based on DNA,    was not programmable and solved a very specific problem in relation with a tour over a set of cities (Adleman, 1994). A \textbf{programmable computer} is one that can implement any recursive function. It is said that a programmable computer has \textbf{universal computation}. Other versions more appropriate for computing science of the equivalence of the universal Turing paradigm and mutation machines      are known since the 1930's: our mutation machines were studied in the form of  \textbf{Post systems} and \textbf{Markov algorithms} and both were proved to be equivalent to Turing machines (Tourlakis, 1984). 
  
  \
  
  Technical detail: computers are composite units, so one can arrange their constitutive parts in different forms to generate a family of architectures. It is therefore mandatory to ask: are all architectures equivalent? No they are not and in fact there is a terrible competence for building the best processors in the industry. But it seems to us that from our stand,   dealing with the general behavior of computers over complex problems, all architectures with a finite number of units are asymptotically equivalent (high levels of complexity make fun of them all). Nevertheless, if one has the opportunity to recruit an infinite number of processors, important differences might appear in relation  with the NP=?P open problem (Clay Mathematics Institute, 2010). All this thematic is very rich and its relation with molecular biology is studied under the trend of seeing computations everywhere in nature (Kari and Rozenberg, 2008). 
  
  \
  
Because we do know that mutation machines are in principle as powerful as modern electronic computers that have allowed us to solve extremely complex problems of design, we conclude that it is worth in science to study the evolutionary theory.

\section{Evolution is scientific}

The evolutionary theory claims  that evolution is the developer responsible for the existence of the software contained in the genome. If one stops here, one is repeating a slogan. It is therefore mandatory to notice that the evolutionary theory has nothing to do with  dogmas: it is science. It is one of the most attractive sciences  because one can work out crucial predictions and challenges to see whether or not they are fulfilled. Our first  duty is to prove that  evolution is better than   chance to solve problems, otherwise evolution would be an unnecessary adorn in fundamental biology.

\subsection{Evolution solves Shakespeare  much better than chance}

Chance is the default explanation of science: chance   can  solve every one problem and therefore it can explain everything. There is yet a trouble:   to solve relatively complex problems and with a significant probability, say, more than 0.05, huge amounts of resources are needed even much  more than those provided by cosmological constraints. Now, science can be understood as an endeavor  to fabricate more reliable options. Example:  

While it is possible to claim that apples fall by chance, science teaches that they fall pulled by gravitational forces. Now, every scientist prefers to believe in gravitational forces and not in chance. Why? The reason is that science  produces predictions that are fulfilled with extreme accuracy.  

 By the same token, while it is possible to claim that chance is the one and only explanation of everything in biology, science teaches that evolution (plus its natural extensions to cover the origin of life and of the universe) is the correct explanation of the existence of species included our own one.
 
 Beliefs must be tested in science. So, one must justify the purpose of inventing evolution: it must produce results faster than chance. To that aim, we make a contest between chance and evolution to solve a concrete  family of problems that depend on a parameter that indicates the complexity of the problem. To fix ideas, let us think of the problem of guessing a phrase with  $n$ letters. This family of problems is called \textbf{Shakespeare} because Dawkins (1986), who  successfully  popularized that problem, took the phrase from Hamlet: \textit{Methinks it is like a weasel},  in reference to the similitude of a  cloud to a  weasel, a carnivorous mammal.
 
 We can witness that chance can solve Shakespeare: we synthesize a string at random and compare it with the goal phrase. If the two coincide, we are done, otherwise we make another try and so on. We say officially that we are executing a \textbf{random walk} in the space of strings.  Looking for a perfect match, we do this for $n=1$ then for $n=2$ and so on until one gets bored. We  notice that chance can indeed solve all instances but   as the number of letters $n$ grows the needed number of trials also grows   as an exponential function of $n$.   We conclude that chance can solve the problem but in an \textbf{exponential regime}.
 
Next, we  program evolution to solve the same family of problems. In fact, the teaching of Darwin  about evolution plus the proposals of Mendel that received plain justification with the discovery of DNA  can be in hindsight summarized in the following form:  evolution   is an automatic  mechanism that enables 1) random mutation leading to variability and 2) accumulation  of those small random changes that are beneficial for surviving and reproduction (Dawkins, l.c., 1986). Thinking of this allows one to program evolution in any language: one devises a population whose members represent plausible solutions to a given problem. They are tested for how good they solve the problem. If some member solves the problem, we are done, else one forms a new generation given preferentially more offspring to the best fitted. One also includes mutation and recombination. The new generation is ready to be tested for fitness. And so on until the problem is solved up to the desired accuracy. A program of this type is called a \textbf{genetic algorithm}. 
 
 Simulations show that evolution solves Shakespeare  but, beware, using a number of trials that is linear in $n$, say, to match a phrase with $n$ letters one needs a number of trials of the form  $n + 100$. We use to say that evolution works in the \textbf{linear regime}, which strongly contrasts with  the exponential one of chance.  This experiment can be carried out in any desktop  in some half hour  if one has the appropriate software installed (see Rodríguez, 2009a). So, many people believe in evolution not because of the scientific propaganda  but because of direct experimentation in which evolution not only overcomes chance but produces results just in front of our eyes. This trend is now the brand of a new generation of scientists.

 \subsection{Evolution is real}
 
Having witnessed how good is evolution to solve Shakespeare, one immediately falls pray of panic because one arrives to the conclusion that evolution is real. The problem with reality is that it is very complex and so one begins to feel that Shakespeare is a game for babies that is weakly related to reality. Which is the fundamental characteristic of reality that possibly could challenge the computing power of evolution? More to the point, can evolution be challenged anyhow?

Yes, complexity can challenge evolution. Why? The reason is that evolution contains no magic: you can implement a recursive function in a DNA computer as well as in any other one. The corresponding program runs in the same regime in all computers, say, if one implementation runs wasting a number of iterations that is polynomial in a given parameter (think of ordering $n$ numbers), then all other implementations in all other paradigms will also  run in the same regime. The same happens for the exponential case.

So, problems can be solved, by evolution or whatever, but it takes time and more complex problems take more time. Now, everyday experience shows that complexity exists, i.e., there are problems whose solution  demand as long times as desired. What does happen if we include this experimental generalization in our study of evolution?

\subsection{The charm of science returns back}

Let us   see how the two visions over evolution, accumulation of small change vs software developing, produce very different predictions if one just keeps in mind that problems can be solved but it takes long time. To fix ideas, let us consider the next  example, which deals with the peaceful death that some persons   experience:

 Humans that  lie in bed because of  a terminal illness sometimes know with a precision of two or three days  that they are going to die and they cease to feel pain, sorrow, hungry or thirst and they say good bye to everybody. For instance, Annie, the 10 years old daughter  of Darwin, said to his father: \textit{I quite thank you} and these are the last words that he remembered from her (Darwin online, 2010).   We must solve two questions. First: How do they know that they are up to die if they never have died before? Second: How could   those aforementioned characteristics had been selected for    given that in normal conditions no dead person leaves progeny?

In the light of the evolutionary theory, the  answer might as simple as the following: by the same token as there exist programmed cell death, apoptosis, there also exists programmed organismic death, which is accompanied with releasing of endorphins and other relaxing hormones and with a communication to the high brain functions and consciousness. Programmed organismic death was awarded by natural selection as a mechanism to control the  population size of people that have no hope.  It is explained in this way how people know that they are going to die. On the other hand, a death in peace is  impressive for us as it was  impressive for ancient people and thus  a trend of selection of the progeny of the dead person that enjoyed those characteristics was created. And so you and me have the possibility to pass away in  inspiring peaceful conditions.

Let us do now an effort to insert our evolutionary explanation of peaceful death into a scientific program: we must produce predictions that could be observed and corroborated by nature. This is done as follows: our premise is that peaceful death appeared by cultural selection of the progeny of  dying persons. Therefore, we must observe what happens when cultural selection has the wrong sign and also when it is absent at all. What do we have here?  Yes: horses also die in peace.  One is always invited to think that this coincidence  of course was caused by the same type of cultural selection as humans went through:   even those horses that were used as war weapons also must go over selection for  controllable wilderness. This possibly includes selection for a peaceful death. So, we find  until here no contradiction between observation and theory.

Upon the aforementioned general predictions of the evolutionary theory, we also have predictions regarding the type of vision that illuminate us. According to the insight  of evolution as accumulation of small change, we can say that any sign of peaceful death was selected and rapidly evolved towards the sublime deaths that we see today. There is nothing to add here.

But if we adopt the vision of evolution as software developer,  we must firmly keep in mind  that everything is eventually possible for evolution but only after many, many trials. In this setting, it is any more plausible to think that selection of a peaceful death could be consistently awarded during long period of time by all cultures: to die among defiant cries  could be    a very valuable  characteristic   for ancient populations pressed by violence, a situation that is very usual among humans. Therefore, we must expect from the members of some human populations to die among cries of war. Where are they? No where:   in the new setting, the  prediction  is false even for warriors.

\

As we see, involving a perception of complexity in  discussions makes them more interesting and the karma of dogmatism is dispensed for. Science is renewed. Another flavour of the relation of complexity with reality  is that natural evolution has been unable of solving up to date some problems:

\subsection{Evolution is a computing machine with limited power }

Evolution   is a computer that relentlessly works to solve the   problem which consists in  calculating well fitted offspring even if the environment changes.  The problem is well posed because the environment includes the offspring of  individuals of other clans which by mere chance might suffer  favorable modifications. Now, one always can imagine desirable changes for any situation. For instance, every human mother would like her children to be more resistant to diseases while they in their turn dream  of  superforce and wings or levitation.   

But the crude reality  is that there is no magic in evolution:  every real computing facility has limited power and problems exist that are unsolvable given some constraints over resources. The first deals with the stability of species and the second with the menace of a sixth mass extinction. Let us analyze the first  problem:

The  fossil record shows that there have been species of bacteria that have  existed in the same form over the last 600 million years (Purves et al, 2001). The mystery is that mutation works everywhere at every moment and so one must clearly understand why species are not permanently shifted in its morphology and physiology. Our explanation is  qualitative:

Every computer needs some time to solve anyone problem. Some problems can be solved very easily while others demand a huge number of operations. The stability of species means that to cause a shift  is at present too a complex problem for the evolutionary computing machine and so it  has not had  enough time to solve it. So, each species in the Earth points to the fact that the computing power of evolution is indeed very limited. 

In regard with the menace  of a sixth   mass extinction of species (Eldredge, 2001), computing science adds: mass extinctions have been common along the history of life on Earth and they are explained by the incapability of the computing evolutionary machine to   solve the problems posed by the   challenges of the  moment.  So, we have every reason to believe that   as the evolutionary machine failed to compute surviving solutions in the past, so it will fail again this time. Therefore, we must convert ourselves from pursuing the next mass extinction to actively pursue conservationist strategies.

\

Let us consider now an example in molecular biology showing that  the evolutionary machine is quite limited in its calculating power:

A \textbf{protein} is a string of amino acids (first structure) that suffers some structural conformation among nearly amino acids (secondary structure), and in the large 3D scale (tertiary structure) and among diverse strings (quaternary structure). How can we assess the similarity among diverse protein structures? A usual methodology is to classify the diverse structures by some descriptors, thank to which proteins are mapped to a given space. Similar structures would appear close one to another. This is the cluster approach to protein structure. A very different approach based on ideal archetypes also has been attempted (Taylor, 2002).

 Some 10 structures might cope for as much as the 30\% of protein structures (Orengo, 1994). This fact might be explained by physical constrains, by   common descent and by sampling effects. On the other hand, it seems that the number of different modes as a protein can fold (in the range of thousands) is larger than the number of folds observed in nature (Chothia, 1992; Taylor,  l.c. 2002). How can we explain that deficit?

Given the eternal persistence of mutation, folding space should be completely occupied. But this is not so: Does   folding space contain bewitched places that somehow could be forbidden to mutation? This question has been answered by   Kuhlman et al (2003), who succeeded in designing a globular protein with a structure that was absent from the data bases.  So, we observe a deficit of occupation  that shall not happen. What do we have to say?

The deficiency in nature of occupancy of folding space shows again that there are (simple) problems that are too heavy for natural evolution. Or, in our terms, the mutation machine used by evolution has limited power, so much  that it is easily overcome by the hardness of the problems that challenge it. 

\

The very fact that the natural  mutation machine has limited power is precisely the reason that enables science, otherwise we would have no more than mythological entities. Anyway, limited as  evolution might be, can it explain the existence of the genome?

\subsection{Genetic programming comes into stage}

If evolution is responsible for  the existence of the genome, we must be able to tame the computing power of evolution to design our own software to produce programs with as   excel quality   as that we see in living creatures from bacteria to a dancing girl. 

It was John Koza (1980) who took seriously the challenge of using evolution as a tool to design software. He showed  with many and diverse  examples that:

\begin{enumerate}
\item The developing of software by means of evolution,  is possible. This enterprise is so generous that it expanded a new discipline that he called \textbf{genetic programming}.
\item  Evolution is better than randomness for the design of software that solve simple problems  (randomness also can develop software, of course, but it is too expensive, maybe wasting more resources than available in the whole universe). 
\item His studies  transmit a  feeling of doubt and intrigue in regard with the power of evolution to overcome randomness for high levels of complexity of genetic programming. Let us, therefore, pay a bit of attention to this relation.
\end{enumerate}

Genetic programming is step by step improving its fame and future (Koza, 2007)

\subsection{ Complexity defies genetic programming }

Our purpose is to elucidate whether or not evolution is always better than chance for hard instances of genetic programming.

Embarking in the study of  a contest between chance and evolution to see which is the best to tailor computer programs means that   one must devise a   \textbf{combinatorial basis} of procedures or methods. To fix ideas, let us think of  a binary adder that must be developed by evolution. To that aim, we form a library of  routines, procedures or modules, each of which  processes a given input, in our example, a string. The initial value of the string encodes for the input, in our example, two binary numbers. Next, the string is recursively elaborated by the procedures of the basis with the purpose of ending  with a string that encodes for the binary addition of the  two numbers. When one combines the elements of that   basis, one gets a program. This can be done by chance or by means of evolution.

When one investigates  chance, one takes a program that has been synthesized at random and tests it: if it fits the function,    one is done, otherwise one generates a new program and so on until success. 

To test evolution, one generates by randomness a set of  programs that compete for fitting the function in the best form. If a given program fits the function, we are done. Otherwise, one picks a sample of programs that include the best ones, those that fit the function in the best approximated form,  recombines and mutates them and gets a new generation that is ready to be tested. And so on until the function is fitted.   This game is full of hard  challenges for the developer and so it is an eternal source of activity, rage and happiness. 

Now, which is better between  chance and evolution for very hard problems? Evolution, of course.  More exactly:

To say that evolution is better than chance in regard with genetic programming means that: 
\begin{enumerate}
\item One can generate  strings  at random by combining the allowed symbols of a programming language, $\{a,b,.. ..,z, 0, 1,..,9, <,>, (,),.. ..\}$ and then test those strings to see whether or not they pass as a program and whether or not they fill in a predefined function. This is feasible for any function but surpassing cosmological constraints.
\item   We always can  devise a  combinatorial basis and  verify in  less than 5 years (time for a thesis) that   evolution can find an approximate solution to any given problem. This pretension is very heavy but to accept it one can think of  those robots, approximations to a human being,  that already walk and think in various labs of the world.  
\end{enumerate}

Let us suppose now that we have a combinatorial basis that can be used by both evolution and chance. Which is better? In all examples that were worked by Koza(l.c., 1980,1996), evolution was better. But those were very simple instances, although of a great variety of cases. What does happen with complex instances of complex problems? It is not known. Anyway, we can see how  intriguing this question is if we  consider the next two  points:

\begin{enumerate}
\item   One can program evolution as bad as desired. This means that it is not rare to invest a significant effort to make a good job but to end with  a program   that behaves worst than randomness even though the solved problem was not that hard (Rodriguez, 2009b).
\item Modelling, explained in the nest  section,  predicts that a high level of complexity always causes evolution to behave worst than chance. So, we can ask:  What is the threshold of complexity above  which evolution behaves worst  than randomness? What is the proportion of evolutionary environments that behave worst than randomness for at least one example? Can one always devise  indeed an evolutionary environment in which evolution could  do it better than randomness and  which is the probability of finding such an environment in nature?
\end{enumerate}

Let us pass now to consider a simple but realistic model of complexity that will be used  to  shade some light on our problems.

\section{SAT is good to model  evolution}

We have used the word complexity in relation with the urgent need to explain our existence using only the available resources that are bounded above by cosmological constraints. For this reason, \textbf{complexity} is a measure of the used resources, specially time, to solve a given problem. The reported output  deals with the number of iterations, processors and bytes of memory that were used to solve a given instance and the form as they vary when instances of the problem are ordered by a given parameter that gauges a sort of  a priori complexity. These thematic is studied by the theory of \textbf{algorithmic complexity}, which is now an extensively developed branch of modern science (Balcázar et al, 1995; Li and Vitányi, 2010). We will benefit ourselves of that theory by rediscovering an example that has been proved to be difficult. 

\

We already know that Shakespeare   can be solved in the linear regime. So, it is very easy. Is this a fair model of evolutionary complexity?

No, it is not. The correct model of evolution is \textbf{SOFT}, the problem of developing software, and this is a problem so complex that it is recommended not to touch at it if one is not armed with lots of expertise: after having too many frustrations, one easily can get the impression that the theory of evolution is devoid of meaning and one might conclude that evolution cannot produce the software of the genome because of scarcity of resources, specially of time. This subjective  appreciation  is very important but needs a revision because of the fact that there are \textbf{hints} that tremendously diminish the complexity of problems. But, what do the hints consist in and how and where do they  operate?

Biological evolution is a software developer as well as genetic programming but there is tremendous difference between them in regard with the level of complexity of the problems that they must solve. The difference  stems from the fact that  biological evolution uses  potent hints while genetic programming has none. The cause is that both types of agents need a combinatorial basis to work upon. A head to head collision with complexity happens when one devises a basis that is well thought but that produces no result after   hours of expectation. One learns in that way that to make a reliable  combinatorial basis is a very hard problem.

The experience of genetic algorithms and genetic programming is that  a reliable combinatorial basis converts a very complex problem in a very simple one in the sense   that it can be solved automatically, by evolution over that basis.  Or, evolution   is a powerful hint that converts problems into automatized algorithms by means of a combinatorial basis. 

We conclude that the complexity of problems is not intrinsic and that the role of hints is determinant. To clearly grasp the power of hints, think of students that make fraud in examinations. By the same token,  natural evolution is armed with hints and so it must solve much easier problems than our corresponding abstract formulations. Henceforth, the subjective and surely authorized feeling that someone can experience over the exceed complexity of biological problems is no guarantee that evolution cannot solve those problems under demanded restrictions. The corresponding question is a scientific problem that must be solved by accepted methodology. To begin with, let us review  two important combinatorial bases that are very important in biology: the immune system and the genetic code.

\

The immune system must calculate solutions to the problem of killing every sort of microbial invaders. To that aim, this system includes at disposition a suitable combinatorial basis to assemble immune globulins (Darnell et al, l.c., 1986). That basis  is tremendously good but not omnipotent, so one person is strong  against certain infectious diseases but weak against others. This machine is subjected to mutation and there is therefore  a tremendous variability among diverse persons: not all members of the same family die at an epidemic. Equivalently, the global immune system (the combinatorial basis of the species) is more potent than separate immune systems: we might die but someone will remain in this world.  Why a person cannot be immune to everything? This deals with another flavor of complexity that is not considered here and that also points to the very limited power of evolution to diffuse into all possible states.

\

Let us see now some details of the hint represented by the genetic code: it establishes   that biological solutions for devising structure and metabolism must come in the form of polypeptides, which are strings composed with 25 amino acids but 5 of them are enzymatic modifications to the 20 fundamental ones that are in the genetic code (Darnell et al, l.c., 1986). For this reason, one uses to think that we need in average around $n^{20}$ trials to synthesize by randomness a polypeptide chain $n$ amino acids long. But evolution is much more efficient. To see this, it is enough to notice that amino acids form families according to diverse physic-chemical criteria, for instance, they are grouped according to whether their side chain are  acidic, basic, uncharged polar or nonpolar (Access Excellence, 2010). Therefore, some variations in within these classes can be classified as of second rank  because they possibly represent changes that weakly modify the properties of the polypeptide and so the problems of replacement might be  easy for evolution. 

If we consider the most severe physic chemical descriptor, which is the relation  between  the side chain of the amino acids and  water, we see only two classes: hydrophobic vs hydrophilic. From this point of view, a polypeptide looks just like this  OOIIIOOOIIIOOI, where O stands for  hydrophObic  while I stands for hydrophIlic. But, what do we have in here? We have at least two possible readings:

 The first is that the hint contained in the genetic code diminishes the complexity of SOFT to that of Shakespeare  for a binary alphabet. We discard  such a reading because it says that we can use evolution to design enzymes for any reaction and that the ensuing problem can be solved during a rainy  evening. We have at present no reason to accept such an idea: evolution is used to design enzymes (Arnold, 2010) but the required expertise is so high  and the ensuing work is so heavy that results  are worth publication in the most famous journals. 
 
 Nevertheless, it is not excluded that  amazing advances in the subject might be achieved in a future, i.e., very powerful hints that diminish the complexity of SOFT to a tractable problem possibly exist. This will cause a tremendous revolution in computing science with such side effects like these: a terrorist group would have the opportunity to design the geometry of a H bomb using a desktop during some few hours. Or a software that effectively breaks down any security system would be sold in the black market  at a price well above   one billion dollars.

A second reading defines our proposal:  we can rewrite  a string of the form OOIIIOOOIIIOOI, that encodes the hydrophilic information of the amino acids of a protein,  in  a slightly different form as TTFFFTTTFFFTTF, where we have changed O by T and I by F. The T comes from \textit{true} and the F from \textit{false}.  Written in this way, a polypeptide looks just like  a plausible  solution to  SAT, the satisfaction problem of mathematical logic, whose aim is to find a satisfying truth  assignment of a logical proposition.

 One can understand what  this problem consists in if we think of the following propositions: $p=$ \textit{somebody worked yesterday},  $q= $  \textit{someone  remained yesterday in the city}. The composite proposition\textit{ somebody worked yesterday and somebody did not remain yesterday in the city} is noted as $ p \wedge \sim q$ and is read as \textit{p and not q}. We say that $ p \wedge \sim q$ is a proposition with 2 binary \textbf{boolean variables} (that can take on two values, false else true,   representing truth values).  This proposition is false for me, because  I did not worked yesterday, and I remained in the city. My behavior yesterday can be encoded as $FT$ because it is false that I worked yesterday and it is true that I remained yesterday in the city. For Natasha the situation is as follows: she was appointed to go to another city to visit some clients of her company and for this reason she left the city yesterday.  So, Natasha functions like TF: she worked yesterday (T) and she traveled, so it is false (F) that she remained in the city.   If we replace $TF$ in $ p \wedge \sim q$, we get $ T \wedge \sim F = T \wedge T = T$.  We say that $TF$ is a satisfying assignment for $ p \wedge \sim q$ because the proposition evaluates to true if we use that  assignment. On the other hand, $TT$ is not a satisfying assignment because with it, the proposition evaluates to false.

\textbf{SAT} is then the problem of finding a satisfying assignment to a proposition if only  that is possible else to declare that no satisfying assignment exists. The problem is well posed because there are propositions that are not satisfiable, say, contradictions, such as \textit{Sulikho left the city but she remained in the city}.

 A proposition could be of the form $(p\vee q) \wedge (\sim p \vee  r)$ that is read as \textit{p or q, and p or not r}. This proposition is said to have 2 \textbf{clauses}, $(p\vee q)$ and $ (\sim p \vee  r)$, and 3 \textbf{binary variables}, $p,q,r$. The complexity of SAT resides in the fact that it correctly captures the contradictory effects that a decision might have on complex systems: good here but bad there, say, if one gives to $p$ the truth value T, that is good for the first clause that evaluates to true but bad for the second one that must wait for a correct truth assignment of $r$ to also evaluate to true.

SAT is ordinary in the human world: it is inspired by language problems in which one must express oneself without falling into contradictions. Besides, we have more than 3000 different interpretations of SAT in problems of life, science and technology (Wikipedia, 2010a). That is why we consider that SAT, with its most difficult instances, is the standard example of ordinary complexity. SAT is   also very rich: some subfamilies of SAT can be solved by an intelligent being at once. Nevertheless, keep in mind that one can program evolution as bad as desired and one may find that some genetic algorithms that solve those obvious problems run in the exponential regime (Rodríguez, 2009c). 

By contrast,  others subfamilies demand,  to the best of our understanding, an exponential regime of work.  If we must find in these cases a satisfying assignment for a given proposition with $n$ boolean variables,  then we must test around $2^n$ assignments to find a satisfying one or to decide that no one can be found. The problem with the exponential regime is that for moderate values of $n$ one must make such a large  number of tests that  the age of the Universe will not be enough to run them if they are done one after the other. But if we use parallelism, the whole universe will not be abundant enough to provide the material to build   the necessary number of processors even if they are molecular units.

The anterior comment is usually compressed as follows: the worst cases of SAT must be solved in the exponential regime. This is experimentally valid for any computing paradigm, be it evolution or not. It is not excluded that this experimental  generalization could be broken, but it seems highly improbable (Bennet, 1981).

\

 How does behave evolution when trying to solve SAT?  One can see that   evolution is very good at finding approximate solutions to SAT, i.e., assignments that satisfy as much as 6/7 of the  clauses of a target proposition are found in front of own eyes (Rodríguez, 2009d). Is this tremendous power a virtue or a pitfall?

\subsection{Half cooked products are important in real life}

Evolution can produce half cooked, approximate  solutions  to SAT almost immediately and the same happens with any other paradigm. Moreover, the same happens with  thousands of problems of the most varied nature. These facts explain our   modern culture in which everyone can learn in a short course the fundamentals of a discipline but cannot get expertise even after 5 years of hard work: just consider the learning of a foreign language or  music. 

Let us try to imagine a situation in which the possibility to find very good approximate solutions to a problem might be converted into a smart idea. Our proposal is to think of  transmission of images, say through Internet. The problem is that images are very informative and so it is very expensive to transmit them. A popular solution consists in dividing the image by a rectangular lattice and to assign one single color to each square. If we are allowed to transmit a small volume of information, we use a very coarse lattice. But if we have no restriction over the volume of transmitted information, we can refine the lattice even to pixels. Let us consider now another approach that defines what must be intelligent transmission, an idea that appeared more than 30 years ago: 

Before sending an image, we try to recognize in it the patterns that it contains, say, faces, mountains, trees.   This is a very complex operation that can be approximated by recognition of simple patterns, say, circles, rectangles, cones, cylinders.   The smart idea consists in going up from simple patterns to more complex to very complex until perfection is achieved. We consider that Internet will collapse if this problem is not quickly solved. 

\

Half cooked products are very important but there are situations in where they create nightmares. The next   case is very important in our deal.

\subsection{Complexity converts evolution in midway  stagnation}

The correct model of evolution is SOFT but we have simplified it to SAT, which can be studied at home,  while we discarded Shakespeare as misleading because of its extreme simplicity. Evolution is quite good to find approximate solutions to SAT but very bad to find exact solutions to its worst instances. From these premises, let us prove that complexity causes evolution to behave worst than randomness:

What does mean that a problem is complex for chance? It means that  one can solve it by randomness but wasting too many trials. Or, a problem is complex when the probability is very low. Chance is  simple and direct: if one  makes a random walk in the space of strings that are plausible solutions, one is done when one finds a string that fulfills the function otherwise one must keep trying.  

What does mean that a solvable problem is complex for evolution? It means that evolution is quite good to find approximate, half cooked solutions of that problem but very bad to find exact, perfect solutions to its worst instances. In those cases, one expects evolution to behave worst  than randomness. Let us try to  mechanistically see  how this happens:

   When  evolution is triggered, the whole world is rapidly  colonized by approximate solutions that are   swiftly substituted by  better ones. As  quality rises, improvement becomes more improbable because it is more probable to change into worst than into better.  Complexity means that stagnation emerges while   products  are anyway half cooked.  Stagnation is compatible with  neutral substitutions, such as it   studied by the theory of Kimura (1993).   Very good solutions exist and indeed can be found by the evolutionary process but to do so one must begin with  worst than the best members of the population (Lenski at al, 2003).  Because they are worst than the fittest, they are selected against. At this stage, complexity means that those unfit individuals from which the solution might descend    appear with very low probability.  That is why one must make new generations with the inclusion of  a sampling procedure that picks up some of the bad  exponents of fathers or to implement hot levels of mutation. Nevertheless, if complexity is high, the probability to hit and  include one that serves is very low because one always work with finite resources. As a consequence, the evolutionary path to perfection is blinded by  selection plus sampling effects. Now,  these factors work at every generation, at every moment. 
   
In hindsight, \textbf{stagnation},  an emerging  systematic delaying effect, appears when complexity is strong enough. That is why the rate of evolution towards a full fledged solution is so retarded that chance could be faster than evolution to solve complex problems.

\

Let us now relate our claims with observable facts.
 
 \subsection{Evolution leaves tracks of half cooked products}

We have seen that evolution is converted by complexity into midway stagnation. In that case, the products that are produced at each period of ecstasy  are classified as half cooked, whose quality has a lot to be desired. The human experience with software design is that one actually goes over a long path filled in half cooked products, that in general are improved over time. That is why we have too many releases. Our arguments show that this phenomenon is mandatory, be it for industry, academy or biological evolution. 

We can now formulate a scientific prediction: \textit{if biological beings appeared by a natural  evolutionary process beginning from some simple creatures that were enabled with evolution, then there must be a path to perfection and complexity which  must be filled in half cooked products}. 

It is tantalizing to say that biology fulfills this prediction:   someone might claim to  see an evolution from bacteria to fishes to amphibians to mammals to chimps to humans in which each intermediate stage corresponds to a period of ecstasy and given the increasing trend in excellence and complexity, a  link can be classified as half cooked with respect to the next. We eagerly deny that such interpretation might be correct. A parable could help to understand what we have in mind:

Let us think of the little girl that someday wanted to  play the piano. How much we enjoyed her improvements everyday as with the piano as with the voice. She went from victory to a higher victory  frequently at the beginning but then she endured  over longer periods of ecstasy.  And sometimes she has mud. How good that she had the tenderness of the mother and the seriousness of the father. She decided  to make a career   as soloist when she was 17. Although she was well prepared, she had a lot to learn and she advanced very slowly: she invested more than a year trying to control the trembling of her knees, two more years were needed for her to insert the first true smile in within her virtuous but cold performance. During the last ten years she has progressed very slowly  in transpiring the passion for life such as great artists do. Anyway, the public and the media take her  as a diva. She marvels at this because she thinks  that she is as yet in the making.
 
 We see in this fairy tail how is the evolution of perfection: half cooked intermediate products abound and persist during long time.  Now, the evolution depicted by evolutionary trees does not fit our description of evolution of perfection  because no one registered being can be given the status of a half cooked product. For instance, the most simple bacterium \emph{Mycoplasma genitalium,} has more than 200 essential enzymes  (Fraser et al, 1995; Mushegian and Koonin, 1996; Hutchison et al, 1999) whose quality of performance is no lower  than those we observe in our organisms. Now, this cannot be explained as the result of evolutionary convergence facilitated by the easiness of the problem of developing an enzyme: if someone can design an enzyme, by whatever method and using whatever hint, he or she is invited to publish his or her results in the most prestigious journals of the planet, such as it indeed happens. 
 
 Thus,   biological evolution lacks the mandatory features characteristic of the evolution of perfection: abundance of half cooked intermediate products.

\

\subsection{The bug law is also valid for natural evolution}

Guided by the human experience, we formulated the bug law of software design:   any developer path always goes over a terrible nightmare of bugs, errors that cause malfunction of  programs and half cooked products, and of corrections that generate more bugs. Our studies of SAT allowed us to rationalize that law in terms of complexity and sampling effects, so we feel free to consider that law as a fact.

 This is the human experience and we wanted to apply it to evolution. The problem is that every piece of software   in the planet exists because it was designed, it was the implementation of a given task that was known and specified beforehand. Therefore, our complexity based law cannot be applied at its face value to the evolutionary theory  because evolution has no aim, no purpose, so it cannot commit bugs, faults against a predefined objective: according to the evolutionary theory, natural evolution   develops software by diffusion, by serendipity,  not but design. Nevertheless, any evolutionary environment   acquires    an emergent  purpose, an emergent status of designer that is intrinsic to  natural selection:

There is no concrete choosing agent in nature that picks the best fitted individuals  but differential surviving plus reproduction can be looked at in hindsight as a picking up activity of an emergent agent that is universally recognized as natural selection. Which is the purpose of this emergent agent? To improve surviving. This is the emergent aim of evolution, of natural selection. Once we have a goal, we also can speak of  errors or bugs against that goal. That is why many people refers to inviable or deleterious mutations as errors. This is allowed in science but one must keep in mind that such a vision is   in hindsight and in regard with the winners of evolution. 

Now that we have managed to unveil a purpose in evolution, we are authorized to apply to it the very same law that is known for humans: any evolutionary path must be   paved with bugs and  includes well differentiated half cooked products. 

This is not expected under the old vision of evolution as accumulation of small change because evolution is  under that vision powerful enough to rapidly find optimal solutions to biological problems. That is why   tracks are so few that they are not expected to appear neither in the fossil record nor in modern extant populations.  But if we adopt the insight of evolution as software developer then    we must firmly keep in mind  that in general the velocity of evolution is very, very slow if any and that many, many trials must be done before a clear trend of evolution could be detected. As a consequence, we expect variability along the fossil record and in modern populations, a variability in which half cooked products dominate.

Technical detail: when we have a problem of design, one must fulfill a very precise function. But in evolution one must not  fulfill an specific function but only to produce something that eventually fits a need under present circumstances. Clearly, design is much more complex than evolutionary developing. Therefore, it is objectionable to try to apply  to evolutionary diffusion those results that are valid for design. This objection is very nice: the complexity of the problem of design is certainly greater than that of developing by diffusion  to see what can result. Nevertheless, both problems are very complex, a fact that is illustrated by a saying: \textit{good luck  strikes only those that are prepared to welcome it}.

\

The best way to proceed now is to try out  a mechanistic discussion. Let us choose a  concrete biological detail, anyone can serve, for instance,   the feathers of aves.  

\section{ Case study: the evolution of feathers}

A bird is an animal   that is dressed in feathers: those at the neck are of middle size and symmetric, those in the tail are   large and symmetric, those of the wing are large, some of which are symmetrical others not.

\subsection{There are 8 types of feathers}

While a non trained eye can distinguish  4 or less  different types of feathers, specialists distinguish 8 (this and all technical details mentioned in this subsection and that appear in italics are taken from The Earthlife Web (2010)). The names of types of feather are: \textit{ remiges, retrices, contour feathers, down feathers, semiplumes, filoplumes, bristles and powder feathers. }

In spite of their variability, feathers have the same  general structure: they have an  axis, which is  a hollow shaft of decreasing diameter, and the \textbf{vane} which consists in \textbf{barbs}. The wide part of the axis that is inserted into the skin is the \textbf{calamus}, while the free part is the  \textbf{rachis}.    In complex feathers, the barbs themselves are also branched and form the \textbf{barbules}, which  have minute hooks called \textbf{barbicels} for cross-attachment (Wikipedia, 2010b).  

\

It is desirable to explain the form and distribution of feathers  by their function. Example:  \textit{bristles have practically no barbs at all and are stiff. They occur around the eyes and mouths of some birds and are protective in function. They are particularly evident in the honey buzzard (Pernis apivorus), which feeds on the nests and young of social bees and wasps and needs protection around its beak from the stings of the adult bees and wasps.} 

On the other hand, we know how to explain the asymmetric form of some feathers:  \textit{a remige [a wing flight feather] .. .. is asymmetrical i.e. the vane is much smaller on one side than the other.  This is because the pressures on the 'leading edge' of the feather (the part that faces forward) are far greater than those on the trailing edge. If the leading edge vane was as large as the trailing edge it would soon become very ragged and not work properly.}

Explanations as that just mentioned are not always available because to understand the form and pattern of distribution of feathers, one must take care of thermodynamics, aerodynamics, material science, structural engineering, physical chemistry  and  ecology to say the least. So, we  believe that the observed in most birds distribution perfectly fits a very complex optimization problem but we as yet cannot prove that. For instance,  \textit{powder feathers occur scattered throughout the plumage of most birds, but their function is not well understood.}  So, complexity inspires an open problem: clearly specify the function that feathers fulfill and then dress the bird in optimal form given a combinatorial basis of 8 types of feathers.  A solution to this question must include an explanation of why  a feather must be so complex, i.e., no one expects the existence of  barbicels until one is acquainted with them. Nevertheless, if  their existence is acknowledged, one can think that this is a super smart solution to the problem of devising an automatically self repairing structure that is    both rigid and breakable.

\

If we think of the form and distribution of feathers in relation with function, we are trying to sustain the a priori idea that things are well done such are they exist. Such a belief is plainly justified, for instance: the resistance of a feather is so high that \textit{the longest feathers in the world belong to an ornamental chicken bread in Japan in 1972, this specimen had tail feathers 10.59m or 34.75ft long.}

\

 But, how did feathers arise? 
 
 \subsection{Feathers appeared with dinosaurs}
 
Modern birds are thought to be descendant of dinosaurs with teeth or at least to have a common ancestor with them (Polin, 1996). Ancient dinosaurs were covered with scales, but some of them exhibited hair in  later periods, which evolved into symmetric feathers and then into asymmetric aerodynamic ones.  Now, this belief is made into a falsifiable theory by the  biogenesis law of evolution (Patten, 1958), which says: a living being repeats along its (embryonic) development the evolutionary history of its species. So, we predict that birds must have teeth, scales and hair in its early embryonic state, next teeth and scales must disappear, and hair must be changed into feathers. 

Now,  if one examines the development of chickens, one sees that they are covered by fine hair when their are born. Some days later, they begin to moult and  hairs are replaced by feathers.  Besides, birds have scales to protect their legs.  But to the best of our knowledge, a chicken never has teeth. Anyway, we consider that the overall pattern of development fits our prediction but only approximately. 

Interlude: we can formulate in our terms the partial fulfillment of our evolutionary predictions: the biogenesis law of evolution is a half cooked product that is very popular  among scientists  and that never will be erased from the history of science even if a perfect reformulation gets unveiled.  Thus, the evolution of science conforms to our complexity based prediction about half cooked products. 

\subsection{The fossil record shows an incremental pattern }

In spite of our protests,  the next  evolutionary theoretical scenario is considered to be stable in regard with feathers (Xu and Guo, 2009):

\begin{enumerate}
\item Birds descended from fishes through dinosaurs.
\item Feathers appeared with dinosaurs and suffered a long period of evolution before birds.
\item The most simple feather was just a single tubular filament.
\item Barbs appeared before than  rachis.
\item  Radially symmetric  plumulaceous feathers are more primitive that bilaterally symmetrical pennaceous feathers.
\item Barbules arrived as the last innovation.
\end{enumerate}

This scenario is based on paleontology, so it is descriptive science, i.e. observations + theory ladeness (Newall, 2005). We see a  simple pattern of increasing complexity that seems to us astonishingly successful. So, it is highly probable to decide that the first feather, a filament, is a  half cooked product as compared with a modern remige. Thus, one can claim that the demands of complexity theory in regard with half cooked products have been fulfilled. Such an interpretation is incorrect. To understand why, let us make a gedanken experiment in which we pay attention to half cooked products.

\subsection{Catching an evolutionary error}

Modern feathers as well as all fossilized ones have their barbs pointing outwards and forward toward the tip of the feather. This of course is very good for flying: under that arrangement, the  wind  will comb the barbs and an aerodynamic silhouette is naturally cared of. 

There is another very technical detail: barbs of modern aves are not straight, they have a curvature whose sign agrees with   aerodynamic requirements. The degree of curvature is well defined and I have the illusion that it is correctly set.  Now, where are the feathers whose barbs point inward, towards the calamus? And, where are the feathers whose curvature goes   in the wrong direction?  And where are the feathers that have no curvature?

 Feathers with curvature in the wrong direction or with no curvature are found in the early stages of evolution (see fig 4 of Xu and Guo, l.c.,  2009). But barbs pointing inward are not found. At this moment, a correction is needed. To say that a curvature has the wrong direction is an expression that has a meaning only for flying beings. But early stages of the evolution of feathers were not guided by flying: heat dissipation and display could be more robust explanations (Xu and Guo, l.c., 2009). Flying appeared as a later use of objects that were selected by other functions. So, we misinterpret evolution if we judge a given type of curvature as wrong. Nevertheless, such assessment is allowed in hindsight and with respect to actual  species. 
 
 All in all, we are left with the question: Where are the feathers whose barbs  point inward, toward the skin? 
 
 In general, a   question of this sort is senseless because no one can pose barriers or conditions to randomness to explain a single, isolated event.  Nevertheless, we have the  duty to impose statistical barriers to randomness when it is used as the explanation of a multitude of independent  events, say more than 10.
 
 Where is a multitude of independent events in relation with feathers? They are here:
 
 We have in first instance that scales and feathers are both done of keratins, a family of proteins.  So, we might think that we have an evolutionary environment with a  combinatorial basis composed of keratins. Given the fossil record, we are invited to propose that in ancient dinosaurs there was a perturbation  of the architectural determinants of keratins  and as a result the evolutionary branch for feathers appeared.  But, what is concretely an architectural perturbation of keratins?
 
A molecular architecture has various determinants (Darnell et al, 1986). We have in first place the amino acid scale: the formation of $\alpha$-helices plus  the interconnection by means of disulphide bridges (whose density is regulated through the proportion of   cysteine, with sulphure at its side chain). Then we have the composition of the ends of the keratins  that determine how ends are tied. At the supramolecular level we have special bridges that are made by specific gluing molecules. We have at last inductors  of conformation at large  that might include groups of cells. On top of all that, we have microscopical units that can be combined, in our case, barbules, barbicels and rachises.
 
One sees that a perturbation of the architecture creates place for  too many  half cooked products, say, if a keratin has  few interchain  disulfide bridges, the material loses    toughness and becomes  flexible but if it has too few of those bridges, the material might become structureless.  Or, imagine a rachis that is appended to the end of a barbicel.

\

Where are all these possible  half cooked products? They are nowhere. Why?
 
 \
 
From the stand of computing science, there is only one answer and it is very simple: the problem of concocting a feather is a tremendously easy problem for evolution and  it was solved so rapidly that the left tracks were not too many and so these had very few opportunities of having appeared in the fossil record. 

The most interesting point is that the same answer serves to respond to all possible scientific questions about evolution. In short, the pretension reads: evolution is magic to solve problems. But such a pretension is any more believable: evolution is today an ordinary tool to solve problems of the most varied nature  and it is perfectly clear that it is not significantly more efficient than other methods or paradigms:   many, many trials of mutation and selection must be done before satisfactory solutions could be accepted. But mechanistically, what does this mean?

\subsection{The thorn bird incarnates true evolution}

Let us  try to imagine what could be the further luck of a feather whose barbs point inwards, to the calamus. Let us see how this type of feather triggered an evolutionary path  that should have been well documented in the fossil record and that must be visible now in modern populations. 

As we know,  keratins are the row material for feathers. These are proteins that are encoded by the DNA and that therefore are subjected to mutation. We also know that one can harden keratins by augmenting the number of disulphide bridges. So, we can envisage a trend of keratin hardening that converts the rachis into a thorn and a rachis with barbs into a harpoon. We end with a thorn bird: it has spines and hooks instead of feathers.

 The immense variability of spines and hooks  is derived from the the immense variability of keratins and the form as they might be combined. Such a variability creates immediately populations with ethological diversity: some use their spines just to play because these are soft. Other use their slightly hardened spines for display and these can be done spectacularly:  feathers are red but their tips are black. Therefore, when  a thorn bird  is in peace, it looks black but when it  feels menaced, it displays its feathers and a thorned red monster appears. In some species the high  hardness of the feathers correlates with the hardness of the beaks and so they are carnivorous. 
 
Most of this variability must be imagined since it never reached our age because of not enough hardening of their spines. But there is one type that should had left  abundant perdurable fossils: one species developed a pair of hooks in the shoulders that could be lost to be reborn again, as nails. These were used as lances to decide leadership in mortal tournaments, as weapons to attack in predation activities and for defense against powerful enemies. One can see a tremendous variability in the fossil record showing variations in the length and thickness of the spines, in its direction and type of hooking. 

And where are the half cooked products? They are here: to have spines instead of feathers creates a serious problem in regard with self damaging. To get non self hurting individuals,  a very long evolutionary game began whose purpose was to find a suitable pattern of  expression of the gene for spines. What we can see in the fossil record as in modern populations is a patchwise expression of the gene for spines that even today is not well stabilized. 

\

Our story of a would-be \textbf{thorn bird}  shows that in evolution there is no clear concept of evolutionary error and that anything can be turned into an advantage. In linguistic terms this reads:\textit{ the meaning is created by the context}.  Nevertheless, it is allowed to say in the evolutionary theory and  in hindsight that a given event was an error with respect to the observed evolutionary path.

 In conclusion, we must expect under the evolutionary theory that the fossil record as well as extant populations to be very rich in evolutionary errors and half cooked products. We consider that this prediction fails: apart from some complaints that are not accepted by everybody, most scientists believe that living beings are marvelous in every sense. 
 
 \subsection{We have discovered the circle}
 
 We have argued  that  the evolutionary theory   can scientifically explain perfection  only if it comes on top of mountains of imperfections. But this is not observed in nature. Quite to the contrary, if one decides to explain life by an  evolutionary process, one must admit that it  seems to be guided by an invisible hand thanks to which the path to perfection to more perfection is guessed at the very first trial. Our complaints are not new. In this regard, it is very instructive to give a look at the version given by Lamarck, who was the first to propose in modern times that species suffer transformations, or that species evolve:
 
 His theory is founded upon two factors (Gould, 2002). First: there is a tendency to acquire complexity beginning form inanimate matter to simple living forms to more complex beings up to man, and, second, this process is modulated  by the interaction of living beings with local circumstances. So, nature functions like a great professor that has a plan for his lecture but that admits with pleasure the questions and interventions of his pupils. As a consequence, the evolution of life followed a path in which half cooked products need not to appear although some exceptions might happen. Actually: what does guide the process along which complexity is improved? In our opinion, this cannot be understood unless one takes into account that Lamarck was a convinced alchemist (Greenberg, 2007) that sees spiritual powers behind matter.

Defying the ideas of Lamarck, Darwin (1859) proposed his theory of evolution by means of natural selection:  complexity and  perfection appeared by accumulation of small change plus differential surviving and reproduction, so, perfection emerges from the struggle for life. The idea was very good but at last it was wrong: in the presence of ordinary complexity, the struggle for life produces too much death of non winning individuals, the half cooked products of evolution, but these are not registered. So, perfection remains there but the evolutionary theory cannot cope with it: the invisible hand of the great professor that serves  as a guide of evolution is not removed. That is why we blame modern evolutionary  theory as disguised lamarckism. 

\

In our words: the modern evolutionary theory is in science a half cooked product. Most probably, there is no greater failure of the evolutionary theory than its incapability to explain why the codons of the genetic code have 3 bases. A genetic code with 4 bases is as natural  as that with three (Neumann et al,  2010) but there is a great difference: with four letters one can encode more than 200 amino acids. Now, 20 well chosen amino acids function better for enzyme design than 200, but 200 amino acids function much better than 20 for diffusion in which there is no goal but only serendipity. So, the evolutionary theory predicts that codons with four bases must invade an evolutionary process which runs on codons with 3 bases: that is not observed in nature. 

 \
 
 There are nevertheless many, many, many, many  facts in favour of the evolutionary theory, which are  summarized in the evolutionary trees.  Our insight of the genome as software allows us to make our own reformulation of this point:

  \section{The genome is evolvable software}
 
When one composes a computer program to fill in a given purpose, one never mind to comply with every detail that the client came with.  Instead, every implementation is a crude approximation to what he or she wants. But more problematic is that the desire of the client   usually mutates every morning and sometimes various times a day. That is why  the first duty of a professional that wants to survive in the deal is to help the client to understand what he or she wants. As a consequence,  every one dreams of evolvable software, an style that eases  changes, amendments, improvements, addition of new functions, evolution toward higher complexity. This is considered to be so important that,  according to modern standards, good software  must be   evolvable. But there is a problem:

 Developing software is very difficult, it is a very harsh activity. Therefore,  reusing  software is mandatory. But reusing  is for the generality of complex tasks  more expensive than ab initio design. This has nothing special, this is simple the general rule for restructuring activities. I learned this from a mason: when I gave to him the task of building a given structure in place of an old one, he began at once  to destroy the old one. So I suggested: It would be cheaper to adequate the old structure.  But he answered: full destruction and a new building is cheaper because  trying to  remaking this old structure into a new one would look as a very delayed destruction. Most software designers learn this lesson by direct experience. So, we must ask: is reusing software competitive for software developing? Let us see: 
 
 No designer is capable of producing software in clean terminated form at his or her first attempt: any developing path is always plagued with every kind of pitfalls.  This can be given in hindsight the next interpretation: every designer reuses half cooked software to produce a better version. So, reuse is immanent to design. Thus, the good idea is to optimize such a process.
 
To begin with, who can imagine a modern programming language without    procedures? These small units are intended to function as tiles for the construction of buildings of the most diverse nature, so they are associated in libraries and APIs (application programming interfaces). The introduction of procedures and subroutines  was the first step   given in the direction of making reusing into a constitutive part of programming. Next we saw the appearing of  parametrized procedures and then of classes.  Then, if we define \textbf{evolvable software} as one that enables reusing,  modification and adaptation to new needs and   during a very long run, then we are saying that \textit{the progress in language development is just the result of a quest for evolvability}, which is a matter of \textbf{architecture}, of \textbf{great design}. Thus, making evolvable software is a profession of its own, which is very tedious:
 
Given a programming language that enables reusing and evolvability, we find that the product of a software designer may be not evolvable, i.e. it would be cheaper to begin a new design rather than trying to adequate it for new needs. But, given that reusability is so important, why most products are not evolvable beyond the needs of the moment? That is a new problem that usually is not solved  because it is not paid. Nevertheless, evolvability at large is possible and appropriate techniques are under research (Mens, 2008). It would be therefore important to see at least one example. Where is it?

\

 The genome is and will remain forever as the most marvelous example of evolvable software: it is so evolvable that its evolvability has been converted into a self contained theory, the evolutionary theory,  which covers the origin of species and of complexity. Now, the evolvability of the genome is what allows modern genetic engineers to produce miracles even though they possibly never have written a single computer program. Where does the evolvability of the genome come form?
 
Evolvability of software is for  human  projects always tied to the effort of a highly qualified   team whose business is big design. By contrast,  evolvability is imagined in the evolutionary theory to   happen autonomously, automatically and circumstantially. We have a problem here:
 
The experience of every software designer is that design that advances thanks to circumstances, to the things that come to the mind while one works, never produces evolvable software. More to the point, no one knows how much one suffers to try to make things with good architectural style but most of times one fails to get even moderate satisfaction (actually, most developers have no time to think of architecture). So, one ends with a   program that resembles a 'big ball of mud', which  'is haphazardly structured, sprawling, sloppy, duct-tape and bailing wire, spaghetti code jungle'(Foote and Yoder, 1999). These properties   suffice to hinder reuse and evolvability. 

It is the same that happens all around the world in developing countries: houses that are built step by step along a period of, say, ten years are structureless, but nevertheless they might  be functional and even pretty, specially if the terrain is not plain. One must contract an architect to get something structured and with good style. Cities obey the same rule: when the city grows enough, one always perceives that streets do not allow a rapid flux of vehicles. Here we need experts in urbanism to remake the city, but  one can observe that modifications at large always are  followed by bigger modifications. In short, we see an evolutionary process that works over  the large scale properties and that is not completely explained by intentional design. So, real human  life intertwines big design and circumstantial change. 

The biological pertinence of these comments can be appreciated at once if we recall that \textit{architecture  is not needed for surviving}:  most software in the world is a spaghetti code jungle and nevertheless software designers are among the best paid professionals. By contrast, very good architecture is needed if one wants evolvable software. This is clear for human projects, but what is observed in biology?

Architectural properties are appreciated  at once in genetics if we recall that  prokaryotes  and eukaryotes differ by the style of design: the genome of the prokaryotes resembles the style that was permitted by ancient programming languages: instructions must go one after another and this is all to  structure. By contrast, the genome of the  eukaryotes resembles programs in modern languages, which allow for highly structured products with units and subunits that are  ready  for use and reuse (Darnell et al, l.c.,  1986), a property that lends itself to evolvability. 

According to  modern biology,    the delicate architectural properties of the genome of eukaryotes arouse circumstantially. In fact, some intermediate stages have been proposed to explain the transition from prokaryotes to eukaryotes. For instance, prokayotes have a simple cellular structure, while eukaryotes have compartmentatilization, say, mithocondria and choloroplasts. This has been explained by assimilation of some prokaryotes  by others, a belief that is partially justified by comparison of DNA sequences (Purves et al, l.c., 2001).   

Nevertheless, these proposals do not explain the change in software  architecture of the genome: good software architecture is not needed to produce structured products.  For instance, if software is perceived by the final user through a GUI (graphic user interface) then  every thing will look extremely ordered. But one always can find behind the GUI a piece of code that in general is very complex and that with certainty is a ball of mud. 

\

The tension between  explanations based on  circumstances vs the need to clearly explain architectural properties is illustrated by the next fact:  the human genome contains more than 100 microbial genes, whose expression is made in the brain, and that do not appear in the genome of the chimp (Whitfield, 2007). This fact announces an architect because those genes that were taken from bacteria are similar to those precious and very expensive stones that are encrusted in strategic sites of palaces to induce a feeling of power and glory, as was done by many kings in ancient times. Today, artworks are preferred. To explain this architectural trait, we have two options:

 If we want an explanation that is  based on circumstances, we simply can invoke horizontal gene transfer that is documented for bacteria (Goldenfeld and Woese, l.c.,  2007) and some eukaryotes (Raymond and   Blankenship, 2003) and postulate its extension form bacteria to chimps and to claim that    mere randomness is all to the origin of man. But one might get inspired by architectural considerations and ask:  What is the probability that a random sampling plus selection would produce such a marvel given that there are millions of species of bacteria (Anissimov, 2010)  and than the number of links from the chimp to man is less than 50 (Stringer, 2010)? This question is killed by modern science as follows: if you cannot explain something by yourself, chance will do it for you:  it is more than enough to say that the requested probability is not zero and with a finite number of discrete units, such as genes, this is always so.

\

We see that science is difficult to understand: it is time to consider the architectural properties of science at large.
  
 \section{Science is regulated by philosophy}

The heart of the  \textbf{scientific method}, such as it is accepted and used everywhere in science, is very simple:  if a fact classifies according to certain model as an extreme event and with relatively very low probability, then it is worth inventing and try a new model to explain it as a normal event. Example:

 If one finds a piece of ice in summer, then we can explain its existence by the default scientific theory: by randomness, i.e. by assuming that the ice arouse by a thermodynamic fluctuation. Nevertheless, such a theory describes  our piece of ice as a very extreme event of an exceedingly low probability. Therefore, we are invited to consider other theories. Of course, we would readily invoke    the existence of refrigerator that describes a piece of ice as a normal event with a necessity of formation.  

We see that the scientific method per se  knows nothing about barriers over inventiveness. Nevertheless, there are some facts about the nature of reality  and of the human being that conspire together to impose  upon science barriers and regulative principles.

\

 The first point about reality is that it is complex and this bears upon the general architecture of formal systems, such as theoretical  science or theology. This is expressed by the  \textbf{theorems of Gödel} (Ebbinghaus et al, 1984; Myers, 2010). The fact is that the   universe is complex enough to generate the following compromise:  if you want coherent non contradictory  thought you must admit mysteries, but if you want to explain everything, you must welcome contradictions.  This means that the connection of science with reality is not and cannot be complete and that science is at last an art as artificial as painting or music: \textit{science and truth are unrelated} and science exists only because we need to tolerate effective models.   Examples:

 We have many sciences: theory of elementary particles, nuclear physics,  thermodynamics, chemistry, molecular biology, biology, ecology, astrophysics. But we have no one single science: Why? It is because we prefer   partial but more or less  consistent descriptions instead of unified but inconsistent pictures. There are wonderful mathematics behind these problems (Primas, 1983).  Most possibly, in no other realm can be seen the artistic nature of science in so clear form as in field theories, which  with their ad hoc  renormalization rules clearly show that we force mathematical thought  to get what we want.

\

In regard with the nature of man, we must take into account that   a human being has no direct access to   \textbf{reality}  if not through his or her  nervous system which produces an \textbf{artificial inner reality} and that is what \textbf{consciousness} sees and perceives. 
The great problem is that the inner artificial reality is  an elaborated simplified description of the reality. For instance, nature has no colors but  only photons, colors are inventions of an ensemble composed of senses, mind and culture (Rodríguez, 2007). 

Since reality is isolated from consciousness, the whole humanity has  been unable to agree on what is real and what is noise. The accepted  solution is to adopt by law a system of regulative principles, a \textbf{philosophy}. Precisely, \textbf{materialism} is the philosophical system that has been accepted by science, according to which the mind is a by-product of matter, which is the true and only reality and as such can be experimentally described and studied with the production of (more or less) consistent results. In response, \textbf{idealism} argues: every scientific experiment is no more than an experiment with the consciousness. So, idealism promotes the importance of the intrinsic uncertainty of the relation of what we perceive and what there exists in reality. 

Idealism enables the existence of \textbf{spiritualism}: there is a reality apart from any mind but a part of it  is only perceived by   consciousness but not by instrumentation. Curiously enough, that reality is always tied to other consciousnesses  that come in the form of spirits, say,  souls of dead persons, angels and demons. If we add to this chart the overwhelming complexity of living beings, we have \textbf{creationism},  the belief that life was created by God, a distinguished spirit which is the only one with the power to create and judge. 

Creationism and evolution are not comparable because they belong in different philosophical systems: creationism is affiliated to idealism while  evolution and its extensions to cover the origin of life and of the universe can be likened to materialism in the sense that we cannot imagine one without the other.   To say that two philosophical systems are not comparable means that a logical reasoning in one system might look as a stupidity in the other: many people prefer to be atheist rather than accept a God that kills innocent people. And conversely, many people regret that  scientists know what happens in the center of the galaxy but ignore that they were created and that they will be judged.  

In spite of all efforts of both evolutionists  and creationists to defeat one another (Editor of Nature, 2008), they remain shoulder to shoulder offering their perspectives to everyone. So it is and so it must be: they produce dissimilar types of mysteries and so the election of one means to choose the type of mysteries and contradictions one wants to survive with. This has implications that hurt   but that clean a lot: 

\begin{itemize}
	\item In regard with science, this means that no scientific falsification is  sufficiently strong to destroy the acceptance of a given theory, as one dreams of when reading the works of Popper (1934). More realistic is the conclusion of Kuhn (1969), according to which the evolution of science is dominated by sociological processes in which new generations fall in love with new insights while those  of the old generation are condemned to be buried with them. 
	\item In regard with theology, the whole picture functions like a tragedy in three acts. Act one: a group of theologians, delighted in their own wisdom, discovers that the fastest way to explain everything is to begin with a contradiction. So, one of them, advances toward the public and says with sublime voice:  \textit{Listen to me, the Almighty will be your defense and your treasure, follow us and shall you have the tenth of the silver of all nations}. Act two: a  dissident   shouts from a balcony unto a group of nuns that go down the street: \textit{Can your Almighty God make a stone that is too heavy for him to carry out?} Act three: the group of theologians expels one man, who walks in three legs and that with weak voice implores: \textit{I  want mercy rather than sacrifices}. 
\end{itemize}

All this happens because it is pretty common to confuse reality with our logical depictions.

\

Dear Reader, if you keep these  clear  concepts in mind, you will make of yourself a blessing and not a curse: you will be able to generate constructive discussions about  the thematic of this article   in spite of whatever  philosophical or religious  affiliation of your  public.

\section{Our insight has a very rich history}

This new insight, that the genome is software and evolution is a software developer, has arrived to us through a rich history. Some points to highlight might be the following:

\begin{enumerate}
\item The Genesis of the Bible,  written probably before 500 BC, declares that the word is so powerful that it was the instrument of God to make his creation. Apostle John     exalted this very idea to such a degree that he declares that  the word and God himself are identical. Words that prescribe precisely specified constructive rules are called  today \textbf{software}. 
\item Everyone uses fingers and pebbles to count. Commerce impulsed the construction of helping machines: the oldest surviving counting board is the Salamis tablet, used by the Babylonians circa 300 B.C., discovered in 1846 on the island of Salamis (ee, 2008).
\item Jean Baptist Lamarck proposes circa 1900 a two factor theory to explain natural order (Gould, l.c.,  2002). The first factor is an architectural plan  that joins simplicity to complexity through an evolutionary process  that includes inheritance of acquired characters. The process is not spontaneous, rather it is guided by alchemical forces that preclude the existence of half cooked products and that circumstantially can create important modulations to the great plan. Such circumstantial modulations conform the second factor. 

\item Charles  Darwin suffers in 1851 the death of his   beloved daughter Anne Elizabeth who was born on  2 March  1841: \textit{oh that she could now know how deeply, how tenderly we do still \& shall ever love her dear joyous face} (Darwin-online, 2010).  While he conserves a faith in life after life and in the power of God to bless the soul of her daughter,  he verifies with this once more that death and automatic   processes have  more explicative power than God. In particular, he proposes that species with their adaptation, coadaptation, complexity and perfection are created by the combined effect of surviving of the fittest in the struggle for life plus  accumulation of small change. He lacks a mechanism to create variability.  (Darwin, 1959).
\item Gregor Mendel-Angustuos   (1865) works with peas to unveil the pattern of inheritance and his studies allow him to   associate inheritance with discrete  units, the genes.  
\item The works of Thomas Hunt Morgan around 1910 over fruit flies (tested by milliards)  allowed him to prove that  genes can mutate. Adding this contributions to that of Mendel, we have now a mechanism to explain both inheritance and variability (Understanding evolution, 2010; Kandel, 2010).
\item  The British mathematician and inventor Charles Babbage worked out in the 19th century the principles of the modern digital computer. Alan Turing and John von Neumann    worked out around 1930 the mathematical fundamentals of computation such as they are known and applied today. Alonzo Church in the same decade began to work on the question of the relation among diverse computing systems. His studies are summarized today in what is known as the Church Thesis:  all computing paradigms are, roughly speaking, mutually  equivalent in regard with efficiency (Tourlakis, l.c., 1984).  A truly digital computer was   built around 1940 by Howard Aiken, a Harvard University mathematician. The ENIAC, Electronic Numerical Integrator And Computer, was built by  John Presper Eckert, Jr. and   John William Mauchly in 1946 (Encarta, 1994).
\item Friedberg (1958) used a blind random search in the space of programs to find a program performing certain task as adding two bits. 
 \item Arthur Samuel (1959) expressed the need of enabling computers with the possibility of learning from experience. From a mechanistic stand point, that task implies to develop an automatic form of devising computer  programs.
 \item The spatial structure of DNA is worked by Watson and Crick (1953) showing that DNA was the perfect candidate to be the carrier of the inheritable genetic information.
 \item Marshal Nirenberg (1961, 1964)    deciphers   the genetic code   showing that the genome is software: the genetic code is an artificial linguistic rule that associates an amino acid to a codon. The proof that this rule is linguistic is that it can be reprogrammed at will: in nature, we have alterations of the genetic code (Gomes et al, 2007). In the industry, we have artificial genetic codes appropriate for the synthesis of new drugs (Neumann et al, l.c.,  2010).
 \item Intel produces in 1971 the first microprocessor, the 4004. It had less than 3000 transistors, a watch frequency of  1MH(Intel, 2010). This humble beginning triggered a  process thanks to which everyone has today the opportunity of testing the evolutionary theory in crucial ways using personal computers. 
 \item John Holland (1975) proves that the theory of evolution encloses a methodology of automatic learning that is  applicable to the    generality of adaptive systems. So, evolution is anymore a patrimony of nature but a tool to be used in ordinary problem solving. A genetic algorithm is a program that uses simulated evolution to solve a given problem of technology, science or art.
 \item John Koza (1980) publishes his \textsl{Genetic programming}. That work showed that the design of software by means of evolution is eventually more of the same evolutionary game of the genetic algorithms. 
 \item  A DNA computer specifically prepared to solve an abstract mathematical problem was built and run by Leonard Adleman (1994).
 \item DNA universal computers (with the possibility of solving every kind of  computable  abstract problems) have been imagined (Beaver, 1996) and it is now clear that diffusion based DNA   universal computation possibly never would reach the industrial production. 
 \item In the 1990's, ARPANET, the Defense Department's inter-computer system, morphed into the National Science Foundation's network and then it was released to the public as   Internet, which enable the transmission of information among computers (Shurkin, 2008).
\item The most crucial point menacing  the security of the communication among computers (a frontal attack upon hosting processors) was  solved in 1995 with the adoption of  the Java virtual machine, a concept developed by Sun Microsystems (Deitel, 2004). The code for Java was  opened  by 2006. Java is important for us because it endows the   scientific community at large with the possibility to make high level computing as in a desktop as in a web. So,  we have the possibility of running Java simulations of evolution on a word wide  scale. Moreover, the modern versions of Java    softly overcomes   C in velocity of execution.  
\end{enumerate}

Most probably, this   vision of the genome as software and of evolution as software developer will appear in the history of science as the contribution of our generation to the understanding of fundamental biology.

\section{Conclusion}

The DNA encodes    instructions that indicate   to the ribosome how to concatenate amino acids to assemble a given protein. The code is   a formal, linguistic rule that can be reprogrammed. For this reason,  we can say that the genome is software and the ribosome is a programmable computer. Now, mutation and recombination can generate new  instructions that eventually can produce a functional protein or enzyme. As a consequence, the evolutionary theory reads: evolution is the software developer responsible for the existence of modern genomes. By its very construction, this reformulation of the evolutionary theory immanently relates evolution with complexity:  the experience of computing science is that software developing is  extremely difficult, an appreciation that is corroborated by direct experience with genetic programming, which uses evolution to develop software,  and by studies of    model problems, say, SAT.  In fact, we have shown for the first time in the history of science that   complexity and evolution have a ring where they can fight at pleasure to give their best.   This new insight    has  been maturing since the 1930's and it  could be both the summary and the most important contribution of our generation to the understanding of evolution.

   \end{document}